\documentclass[aps,preprint,nofootinbib,floatfix]{revtex4}%
\usepackage{amssymb}
\usepackage{amsmath}
\usepackage{epsfig}
\usepackage{amsfonts}
\usepackage{graphicx}%
\providecommand{\U}[1]{\protect\rule{.1in}{.1in}}
\begin{document}
\title{Recurrence relations of Kummer functions and Regge string scattering amplitudes}
\author{Jen-Chi Lee}
\email{jcclee@cc.nctu.edu.tw}
\affiliation{Department of Electrophysics, National Chiao-Tung University, Hsinchu, Taiwan, R.O.C.}
\affiliation{National Center for Theoretical Sciences, Hsinchu, Taiwan, R.O.C.}
\affiliation{}
\author{Yoshihiro Mitsuka}
\email{yoshihiro.mitsuka@gmail.com}
\affiliation{Department of Electrophysics, National Chiao-Tung University, Hsinchu, Taiwan, R.O.C.}
\affiliation{National Center for Theoretical Sciences, Hsinchu, Taiwan, R.O.C.}

\begin{abstract}
We discover an infinite number of recurrence relations among Regge string
scattering amplitudes \cite{bosonic,RRsusy} of different string states at
arbitrary mass levels in the open bosonic string theory. As a result, all
Regge string scattering amplitudes can be algebraically solved up to
multiplicative factors. Instead of decoupling zero-norm states in the fixed
angle regime, the calculation is based on recurrence relations and addition
theorem of Kummer functions of the second kind. These recurrence relations
among Regge string scattering amplitudes are dual to linear relations or
symmetries among high-energy fixed angle string scattering amplitudes
discovered previously.

\end{abstract}
\maketitle

\bigskip%
\setcounter{equation}{0}
\renewcommand{\theequation}{\arabic{section}.\arabic{equation}}%

\section{Introduction}

There are two high-energy regimes of string scattering amplitudes, the fixed
angle regime or Gross Regime (GR) \cite{GM,GrossManes}, and the fixed momentum
transfer regime or Regge regime (RR) \cite{RR1,RR2,RR3,RR4,RR5,RR6,RR7}. The
scattering amplitudes of these two regimes contain information complementary
to each other \cite{bosonic}. Recently a saddle point method was used to
calculate string-tree amplitudes in the GR for string states at arbitrary mass
levels \cite{ChanLee1,ChanLee2,PRL,CHLTY}. See also the developments in
\cite{Sagnotti,West1,West2,Moore}. The ratios of scattering amplitudes among
different string states at each fixed mass level in the GR can be extracted
and were found to be independent of scattering energy and scattering angle.
Alternatively, these infinite number of ratios can be rederived algebraically
by solving linear relations or stringy Ward identities based on the
calculation of decoupling of zero-norm states (ZNS) \cite{ZNS1,ZNS3,ZNS2} in
the string spectrum. These infinite number of ratios were interpreted as
high-energy spacetime symmetry of string theory conjectured by Gross more than
twenty years ago \cite{Gross}.

In contrast to the GR, things were less known in the RR. It was known that
there is no saddle point method for the calculation of Regge string scattering
amplitudes. More recently, a subset of Regge string scattering amplitudes were
calculated by a direct method and it was found that all the Regge string
scattering amplitudes can be expressed in terms of Kummer functions of the
second kind \cite{bosonic,RRsusy}. Moreover, these Regge string scattering
amplitudes can be used to reextract the ratios among high-energy fixed angle
scattering amplitudes calculated in the GR mentioned above \cite{LYAM,MK}.
Presumably, there must be intimate link between string scatterings in the GR
and RR. Unlike the case of the GR, the relations among Regge string scattering
amplitudes of different string states were not understood and the ratios were
suspected to depend on the scattering energy and scattering angle. A natural
question is then raised:

\bigskip

Are there any kind of relations among Regge string scattering amplitudes of
different string states which are dual to those linear relations
\cite{ChanLee1,ChanLee2,PRL,CHLTY} calculated in the fixed angle regime?

\bigskip

To answer this question, in this paper we first calculate the complete set of
Regge string scattering amplitudes which turns out to be much more numerous
than those in the GR. We consider four-point function with three tachyons and
one most general Regge string state. Note that, in order to calculate Regge
stringy Ward identities based on decoupling ZNS, one needs to use the complete
Regge string scattering amplitudes. We then derive Regge stringy Ward
identities for the first few mass levels based on the decoupling of ZNS. We
found that, unlike the case for the GR, the Regge stringy Ward identities were
not good enough to solve all the Regge string scattering amplitudes
algebraically. This is due to the much more numerous Regge string scattering
amplitudes than those in the GR at each fixed mass level. However, we found
that, for mass levels $M^{2}=2$ and$\ 4$, all the Regge stringy Ward
identities can be explicitly proved by using a set of identities of Kummer
function, namely, the recurrence relations of Kummer functions of the second
kind \cite{Slater}. Presumably, the calculation can be generalized to
arbitrary higher mass levels. We then went a step further and show that,
instead of Regge stringy Ward identities, one can use these recurrence
relations of Kummer functions to solve all Regge string scattering amplitudes
at arbitrary higher mass levels algebraically up to multiplicative factors.

Finally, we calculate explicitly some examples of recurrence relations among
Regge string scattering amplitudes based on recurrence relations of Kummer
functions. These recurrence relations are, in general, independent of Regge
stringy Ward identities calculated from decoupling of ZNS. They are thus more
general than Regge stringy Ward identities. In addition, as an example, we
construct an inter-mass level recurrence relation which was calculated by
using addition theorem of Kummer functions \cite{Slater}. Following the same
procedure, one can construct infinite number of recurrence relations among
Regge string scattering amplitudes at arbitrary mass levels. The recurrence
relations for Regge string scattering amplitudes are reminiscent of BCJ
relations \cite{BCJ1,BCJ2,BCJ3,BCJ4,BCJ5} for Yang-Mills gluon color-stripped
scattering amplitudes where kinematic variables show up in the coefficients of
the relations.

\section{GR and RR String Amplitudes}

We begin with a brief review of high-energy string scattering in the fixed
angle regime,%
\begin{equation}
s,-t\rightarrow\infty,t/s\approx-\sin^{2}\frac{\phi}{2}=\text{fixed (but }%
\phi\neq0\text{)} \label{1}%
\end{equation}
where $s=-(k_{1}+k_{2})^{2},$ $t=-(k_{2}+k_{3})^{2}$ and $u=-(k_{1}+k_{3}%
)^{2}$ are the Mandelstam variables and $\phi$ is the center of mass
scattering angle. It was shown \cite{PRL,CHLTY} that for the 26D open bosonic
string the only states that will survive the high-energy limit at mass level
$M_{2}^{2}=2(N-1)$ are of the form (we choose the second state of the
four-point function to be the higher spin string state)
\begin{equation}
\left\vert N,2m,q\right\rangle \equiv(\alpha_{-1}^{T})^{N-2m-2q}(\alpha
_{-1}^{L})^{2m}(\alpha_{-2}^{L})^{q}|0,k\rangle, \label{2}%
\end{equation}
where the polarizations of the 2nd particle with momentum $k_{2}$ on the
scattering plane were defined to be $e^{P}=\frac{1}{M_{2}}(E_{2}%
,\mathrm{k}_{2},0)=\frac{k_{2}}{M_{2}}$ as the momentum polarization,
$e^{L}=\frac{1}{M_{2}}(\mathrm{k}_{2},E_{2},0)$ the longitudinal polarization
and $e^{T}=(0,0,1)$ the transverse polarization which lies on the scattering
plane. $\eta_{\mu\nu}=diag(-1,1,1).$ In Eq.(\ref{2}), $N,m$ and $q$ are
non-negative integers and $N\geq2m+2q.$ These integers characterise the mass
square and "spin" of the higher string states. Note that $e^{P}$ approaches to
$e^{L}$ in the GR \cite{ChanLee2}. So we did not put $e^{P}$ components in
Eq.(\ref{2}). For simplicity, we choose the particles associated with momenta
$k_{1}$, $k_{3}$ and $k_{4}$ to be tachyons. It turned out that the
high-energy fixed angle scattering amplitudes can be calculated by using the
saddle-point method \cite{PRL} to be
\begin{equation}
\mathcal{T}^{(N,2m,q)}=\sqrt{\frac{2\pi}{Kf_{0}^{\prime\prime}}}e^{-Kf_{0}%
}\left[  (-1)^{N-q}\frac{2^{N-q-2m}(2m)!}{m!\ {M}_{2}^{q+2m}}\ \tau^{-\frac
{N}{2}}(1-\tau)^{\frac{3N}{2}}E^{N}+O(E^{N-2})\right]  \label{GRAM}%
\end{equation}
where $K\equiv-k_{1}.k_{2}\rightarrow2E^{2},$ $f(x)\equiv\ln x-\tau\ln(1-x)$
$,\tau\equiv-\frac{k_{2}.k_{3}}{k_{1}.k_{2}}\rightarrow\sin^{2}\frac{\phi}{2}%
$, and the saddle-point for the integration of moduli, $x=x_{0}=\frac
{1}{1-\tau}$, is defined by $f^{\prime}(x_{0})=0$. The complete ratios among
high-energy fixed angle string scattering amplitudes of different string
states at each fixed mass level can be easily calculated from Eq.(\ref{GRAM})
to be \cite{PRL}%
\begin{equation}
\frac{T^{(N,2m,q)}}{T^{(N,0,0)}}=\left(  -\frac{1}{M_{2}}\right)
^{2m+q}\left(  \frac{1}{2}\right)  ^{m+q}(2m-1)!!. \label{3}%
\end{equation}

Alternatively, it was important to discover that \cite{PRL,CHLTY} the ratios
can be calculated by the method of decoupling of two types of ZNS
\begin{equation}
\text{Type I}:L_{-1}\left\vert x\right\rangle ,\text{ where }L_{1}\left\vert
x\right\rangle =L_{2}\left\vert x\right\rangle =0,\text{ }L_{0}\left\vert
x\right\rangle =0;
\end{equation}

\begin{equation}
\text{Type II}:(L_{-2}+\frac{3}{2}L_{-1}^{2})\left\vert \widetilde{x}%
\right\rangle ,\text{ where }L_{1}\left\vert \widetilde{x}\right\rangle
=L_{2}\left\vert \widetilde{x}\right\rangle =0,\text{ }(L_{0}+1)\left\vert
\widetilde{x}\right\rangle =0.
\end{equation}
in the old covariant first quantized string spectrum. The key step was to use
the decoupling of high-energy fixed angle ZNS
\begin{align}
L_{-1}\left\vert n-1,2m-1,q\right\rangle  &  \simeq M\left\vert
n,2m,q\right\rangle +(2m-1)\left\vert n,2m-2,q+1\right\rangle ,\label{HZNS1}\\
L_{-2}\left\vert n-2,0,q\right\rangle  &  \simeq\frac{1}{2}\left\vert
n,0,q\right\rangle +M\left\vert n,0,q+1\right\rangle \label{HZNS2}%
\end{align}
to deduce the ratios of all amplitudes at the leading order energy in
Eq.(\ref{3}). Since the decoupling of ZNS applies to all string loop order,
the ratios calculated in Eq.(\ref{3}) were valid to all string loop order and
were interpreted as high-energy spacetime symmetry of string theory
conjectured by Gross in 1988 \cite{Gross}.

Another high-energy regime of string scattering amplitudes, which contains
dual information of the theory, is the fixed momentum transfer regime or RR.
That is in the kinematic regime%
\begin{equation}
s\rightarrow\infty,-t=\text{fixed (but }-t\neq\infty).
\end{equation}
It was found \cite{bosonic} that the number of high energy scattering
amplitudes for each fixed mass level in this regime is much more numerous than
that of fixed angle regime calculated previously. The complete leading order
high-energy open string states in the Regge regime at each fixed mass level
$N=\sum_{n,m,l>0}np_{n}+mq_{m}+lr_{l}$ are%
\begin{equation}
\left\vert p_{n},q_{m},r_{l}\right\rangle =\prod_{n>0}(\alpha_{-n}^{T}%
)^{p_{n}}\prod_{m>0}(\alpha_{-m}^{P})^{q_{m}}\prod_{l>0}(\alpha_{-l}%
^{L})^{r_{l}}|0,k\rangle. \label{RR}%
\end{equation}
The case for $q_{m}=0$ has been calculated previously in \cite{bosonic,RRsusy}%
. We stress that the inclusion of both $\alpha_{-m}^{P}$ and $\alpha_{-l}^{L}$
operators in Eq.(\ref{RR}) will be crucial to study Regge string Ward
identities to be discussed in the later part of the paper. The momenta of the
four particles on the scattering plane are%

\begin{align}
k_{1}  &  =\left(  +\sqrt{p^{2}+M_{1}^{2}},-p,0\right)  ,\\
k_{2}  &  =\left(  +\sqrt{p^{2}+M_{2}^{2}},+p,0\right)  ,\\
k_{3}  &  =\left(  -\sqrt{q^{2}+M_{3}^{2}},-q\cos\phi,-q\sin\phi\right)  ,\\
k_{4}  &  =\left(  -\sqrt{q^{2}+M_{4}^{2}},+q\cos\phi,+q\sin\phi\right)
\end{align}
where $p\equiv\left\vert \mathrm{\vec{p}}\right\vert $, $q\equiv\left\vert
\mathrm{\vec{q}}\right\vert $ and $k_{i}^{2}=-M_{i}^{2}$. The relevant
kinematics are%
\begin{equation}
e^{P}\cdot k_{1}\simeq-\frac{s}{2M_{2}},\text{ \ }e^{P}\cdot k_{3}\simeq
-\frac{\tilde{t}}{2M_{2}}=-\frac{t-M_{2}^{2}-M_{3}^{2}}{2M_{2}};
\end{equation}%
\begin{equation}
e^{L}\cdot k_{1}\simeq-\frac{s}{2M_{2}},\text{ \ }e^{L}\cdot k_{3}\simeq
-\frac{\tilde{t}^{\prime}}{2M_{2}}=-\frac{t+M_{2}^{2}-M_{3}^{2}}{2M_{2}};
\end{equation}
and%
\begin{equation}
e^{T}\cdot k_{1}=0\text{, \ \ }e^{T}\cdot k_{3}\simeq-\sqrt{-{t}}%
\end{equation}
where $\tilde{t}$ and $\tilde{t}^{\prime}$ are related to $t$ by finite mass
square terms%
\begin{equation}
\tilde{t}=t-M_{2}^{2}-M_{3}^{2}\text{ , \ }\tilde{t}^{\prime}=t+M_{2}%
^{2}-M_{3}^{2}.
\end{equation}
Note that, unlike the case of GR, here $e^{P}$ does not approach to $e^{L}$ in
the RR. The Regge string scattering amplitudes can then be explicitly
calculated to be%
\begin{align}
A\left(  s,t\right)   &  \simeq\int_{0}^{1}dy\ y^{k_{1}k_{2}}(1-y)^{k_{2}%
k_{3}}\cdot\prod_{n}\left[  -\frac{\left(  n-1\right)  !e^{T}\cdot k_{1}%
}{\left(  -y\right)  {}^{n}}-\frac{\left(  n-1\right)  !e^{T}\cdot k_{3}%
}{\left(  1-y\right)  ^{n}}\right]  ^{p_{n}}\nonumber\\
&  \quad\cdot\prod_{m}\left[  \frac{\left(  m-1\right)  !e^{P}\cdot k_{1}%
}{\left(  -y\right)  ^{m}}+\frac{\left(  m-1\right)  !e^{P}\cdot k_{3}%
}{\left(  1-y\right)  ^{m}}\right]  ^{q_{m}}\nonumber\\
&  \cdot\prod_{l}\left[  -\frac{\left(  l-1\right)  !e^{L}\cdot k_{1}}{\left(
-y\right)  ^{l}}-\frac{\left(  l-1\right)  !e^{L}\cdot k_{3}}{\left(
1-y\right)  ^{l}}\right]  ^{r_{l}}\nonumber\\
&  \approx\int_{0}^{1}dy\ y^{-\frac{s}{2}+N-2}\left(  1-y\right)  ^{-\frac
{t}{2}+N-2}\nonumber\\
&  \quad\cdot\prod_{n>0}\left[  \frac{\left(  n-1\right)  !\sqrt{-t}}{\left(
1-y\right)  ^{n}}\right]  ^{p_{n}}\cdot\prod_{m>1}\left[  -\frac{\left(
m-1\right)  !\frac{\tilde{t}}{2M_{2}}}{\left(  1-y\right)  ^{m}}\right]
^{q_{m}}\cdot\prod_{l>1}\left[  \frac{\left(  l-1\right)  !\frac{\tilde
{t}^{\prime}}{2M_{2}}}{\left(  1-y\right)  ^{l}}\right]  ^{r_{l}}\nonumber\\
&  \quad\cdot\left[  \frac{\frac{s}{2M_{2}}}{y}-\frac{\frac{\tilde{t}}{2M_{2}%
}}{\left(  1-y\right)  }\right]  ^{q_{1}}\left[  -\frac{\frac{s}{2M_{2}}}%
{y}+\frac{\frac{\tilde{t}^{\prime}}{2M_{2}}}{\left(  1-y\right)  }\right]
^{r_{1}}\nonumber\\
&  =\prod_{n>0}\left[  \left(  n-1\right)  !\sqrt{-t}\right]  ^{p_{n}}%
\cdot\prod_{m>1}\left[  -\left(  m-1\right)  !\frac{\tilde{t}}{2M_{2}}\right]
^{q_{m}}\cdot\prod_{l>1}\left[  \left(  l-1\right)  !\frac{\tilde{t}^{\prime}%
}{2M_{2}}\right]  ^{r_{l}}\nonumber\\
&  \quad\cdot\sum_{i,j}\binom{q_{1}}{i}\binom{r_{1}}{j}\left(  -\frac
{s}{\tilde{t}}\right)  ^{i}\left(  -\frac{s}{\tilde{t}^{\prime}}\right)
^{j}B\left(  -\frac{s}{2}+N-1-i-j,-\frac{t}{2}-1+i+j\right)  . \label{beta}%
\end{align}
In the second equality of the above equation, we have dropped the first term
in the bracket with power of $p_{n}$, and the first terms in the brackets with
powers of $q_{m}$ and $r_{l}$ for $m,l>1.$ These terms lead to subleading
order terms in energy in the Regge limit \cite{bosonic,RRsusy}. Now the beta
function in Eq.(\ref{beta}) can be approximated in the RR by
\cite{bosonic,RRsusy}
\begin{equation}
B\left(  -\frac{s}{2}+N-1-i-j,-\frac{t}{2}-1+i+j\right)  =B\left(  -\frac
{s}{2}-1,-\frac{t}{2}-1\right)  \left(  -\frac{s}{2}\right)  ^{-i-j}\left(
-\frac{t}{2}-1\right)  _{i+j}%
\end{equation}
where $(a)_{j}=a(a+1)(a+2)...(a+j-1)$ is the Pochhammer symbol. Finally we
arrive at the amplitude with two equivalent expressions
\begin{align}
A\left(  s,t\right)   &  =\prod_{n>0}\left[  \left(  n-1\right)  !\sqrt
{-t}\right]  ^{p_{n}}\cdot\prod_{m>0}\left[  -\left(  m-1\right)
!\frac{\tilde{t}}{2M}\right]  ^{q_{m}}\cdot\prod_{l>1}\left[  \left(
l-1\right)  !\frac{\tilde{t}^{\prime}}{2M}\right]  ^{r_{l}}\label{factor1}\\
&  \quad\cdot B\left(  -\frac{s}{2}-1,-\frac{t}{2}+1\right)  \left(  \frac
{1}{M}\right)  ^{r_{1}}\nonumber\\
&  \cdot\sum_{i=0}^{q_{1}}\binom{q_{1}}{i}\left(  \frac{2}{\tilde{t}}\right)
^{i}\left(  -\frac{t}{2}-1\right)  _{i}U\left(  -r_{1},\frac{t}{2}%
+2-i-r_{1},\frac{\tilde{t}^{\prime}}{2}\right) \nonumber\\
&  =\prod_{n>0}\left[  \left(  n-1\right)  !\sqrt{-t}\right]  ^{p_{n}}%
\cdot\prod_{m>1}\left[  -\left(  m-1\right)  !\frac{\tilde{t}}{2M}\right]
^{q_{m}}\cdot\prod_{l>0}\left[  \left(  l-1\right)  !\frac{\tilde{t}^{\prime}%
}{2M}\right]  ^{r_{l}}\label{factor2}\\
&  \cdot B\left(  -\frac{s}{2}-1,-\frac{t}{2}+1\right)  \left(  -\frac{1}%
{M}\right)  ^{q_{1}}\nonumber\\
&  \cdot\sum_{j=0}^{r_{1}}\binom{r_{1}}{j}\left(  \frac{2}{\tilde{t}^{\prime}%
}\right)  ^{j}\left(  -\frac{t}{2}-1\right)  _{j}U\left(  -q_{1},\frac{t}%
{2}+2-j-q_{1},\frac{\tilde{t}}{2}\right)  .\nonumber
\end{align}
$U$ in Eqs.(\ref{factor1}) and (\ref{factor2}) is the Kummer function of the
second kind and is defined to be%
\begin{equation}
U(a,c,x)=\frac{\pi}{\sin\pi c}\left[  \frac{M(a,c,x)}{(a-c)!(c-1)!}%
-\frac{x^{1-c}M(a+1-c,2-c,x)}{(a-1)!(1-c)!}\right]  \text{ \ }(c\neq2,3,4...)
\label{10}%
\end{equation}
where $M(a,c,x)=\sum_{j=0}^{\infty}\frac{(a)_{j}}{(c)_{j}}\frac{x^{j}}{j!}$ is
the Kummer function of the first kind. $U$ and $M$ are the two solutions of
the Kummer equation%
\begin{equation}
xy^{^{\prime\prime}}(x)+(c-x)y^{\prime}(x)-ay(x)=0. \label{11}%
\end{equation}
It is interesting to note that the Regge behavior is universal and is mass
level independent \cite{bosonic}%
\begin{equation}
B\left(  -1-\frac{s}{2},-1-\frac{t}{2}\right)  \sim s^{\alpha(t)}\text{ \ (in
the RR)} \label{power}%
\end{equation}
where $\alpha(t)=\alpha(0)+\alpha^{\prime}t$, \ $\alpha(0)=1$ and
$\alpha^{\prime}=1/2.$ That is, the well known $\sim s^{\alpha(t)}$ power-law
behavior of the four tachyon string scattering amplitude in the RR can be
extended to arbitrary higher string states. This result will be used to
construct an inter-mass level recurrence relation for Regge string scattering
amplitudes later in Eq.(\ref{RRT3}).

At this point, it is crucial to note that, in our case of Eq.(\ref{factor1})
and Eq.(\ref{factor2}), $c=c(t)$ and is not a constant as in the usual
definition, so $U$ in the Regge string amplitudes is\textit{ not} a solution
of the Kummer equation. This will make some analysis more complicated On the
contrary, since $a=-q_{1}($or $-r_{1})$ a nonpositive integer, the Kummer
functions in Eq.(\ref{factor1}) and Eq.(\ref{factor2}) terminated to a finite
sum. This will simplify the manipulation of Kummer functions used in this
paper. For example, since $c=c(t)$ is not a constant, derivative relations of
Kummer functions in Eq.(\ref{DE1}) to Eq.(\ref{DE6}) are no longer true.
However, the recurrence relations in Eq.(\ref{RC1}) to Eq.(\ref{RC6}) are
still valid as they can be easily verified for the case of finite number of
terms. There is another form of Kummer function and is given by
\begin{align}
U(a,c,x)  &  =\frac{1}{x^{a}}{}_{2}F_{0}\left(  a,1+a-c,-\frac{1}{x}\right)
\nonumber\\
&  \equiv\frac{1}{x^{a}}\sum_{k=0}^{\infty}\frac{1}{k!}\left(  a\right)
_{k}\left(  1+a-c\right)  _{k}\left(  -\frac{1}{x}\right)  ^{k}\ .
\label{finite}%
\end{align}
It is easy to see that the summation in Eq.(\ref{finite}) terminates to a
finite sum for a nonpositive integer $a.$

\section{Recurrence Relations and RR Stringy Ward Identities}

In this section, we first discuss Regge stringy Ward identities derived from
Regge zero-norm states for mass level $M^{2}=2$ and$\ 4.$ We will see that,
unlike the case for the GR we did in Eq.(\ref{HZNS1}) and Eq.(\ref{HZNS2}),
Regge string Ward identities are not good enough to solve all Regge string
scattering amplitudes algebraically. On the other hand, we found that the
recurrence relations of Kummer functions Eq.(\ref{RC1}) to Eq.(\ref{RC6})
discussed in the appendix can be used to prove all Regge stringy Ward
identities. Presumably the calculation can be generalized to arbitrary mass
levels. Another reason to work on recurrence relations of Kummer functions
instead of \ Regge stringy Ward identities is that the former is very easy to
generalize to arbitrary higher mass levels while the latter is not.

Most importantly, for Kummer functions $U(a,c,x)$ in Regge string amplitudes
in Eq.(\ref{factor1}) and Eq.(\ref{factor2}) with $a=-q_{1}($or $-r_{1})$ a
nonpositive integer, one can use recurrence relations to solve all
$U(-q_{1},c,x)$ functions algebraically and thus determine all Regge string
scattering amplitudes at arbitrary mass levels algebraically up to
multiplicative factors. We stress that for general values of $a$, the best one
can obtain from recurrence relations is to express any Kummer function in
terms of any two of its associated function (see the appendix).

There are $9$ Regge string amplitudes for the mass level $M^{2}=2$,
$T^{PP}(\alpha_{-1}^{P}\alpha_{-1}^{P})$,$\ T^{PL}(\alpha_{-1}^{P}\alpha
_{-1}^{L})$, $T^{PT}(\alpha_{-1}^{P}\alpha_{-1}^{T})$, $T^{LL}(\alpha_{-1}%
^{T}\alpha_{-1}^{T})$, $T^{LT}(\alpha_{-1}^{L}\alpha_{-1}^{T})$,
$T^{TT}(\alpha_{-1}^{T}\alpha_{-1}^{T})$, $T^{P}(\alpha_{-2}^{P})$,
$T^{L}(\alpha_{-2}^{L})$, $T^{T}(\alpha_{-2}^{T}).$ For this mass level
$\tilde{t}=t,$ $\tilde{t}^{\prime}=t+4.$ The Regge zero-norm states (RZNS) in
Eq.(\ref{R2.1}) and Eq.(\ref{R2.2}) gives two Regge stringy Ward identities%
\begin{equation}
T^{T}-\sqrt{2}T^{PT}=0, \label{W2.1}%
\end{equation}%
\begin{equation}
T^{L}-\sqrt{2}T^{PL}=0. \label{W2.2}%
\end{equation}
The RZNS in Eq.(\ref{R2.3}) gives%
\begin{equation}
\sqrt{2}T^{P}-T^{PP}-\frac{1}{5}T^{LL}-\frac{1}{5}T^{TT}=0. \label{W2.3}%
\end{equation}
It's obvious to see that these three Regge stringy Ward identities
Eq.(\ref{R2.1}) to Eq.(\ref{R2.3}) are not good enough to solve all the $9$
Regge string scattering amplitudes algebraically. Indeed, the amplitude
$T^{LT}$ does not even show up in any of these three Ward identities.

Instead of Regge stringy Ward identities, in the following we will do the
calculation based on recurrence relations of Kummer functions. We want to
prove these three Regge stringy Ward identities by using recurrence relations%
\begin{align}
U(a-1,c,x)-(2a-c+x)U(a,c,x)+a(1+a-c)U(a+1,c,x)  &  =0,\label{A1}\\
U(a,c,x)-aU(a+1,c,x)-U(a,c-1,x)  &  =0. \label{A2}%
\end{align}
First, by taking some special values of arguments of Kummer function in
Eq.(\ref{A1}) and Eq.(\ref{A2}), one easily obtain
\begin{equation}
U\left(  -1,x,x\right)  =0, \label{R1}%
\end{equation}%
\begin{equation}
U\left(  -2,x,x\right)  +xU(0,x,x)=0 \label{R2}%
\end{equation}
and%

\begin{equation}
U\left(  0,c,x\right)  -U(0,c-1,x)=0. \label{R3}%
\end{equation}
By using Eq.(\ref{factor2}), one easily see that the Ward identity
Eq.(\ref{W2.1}) implies
\begin{equation}
U\left(  0,\frac{t}{2}+2,\frac{\tilde{t}}{2}\right)  +U\left(  -1,\frac{t}%
{2}+1,\frac{\tilde{t}}{2}\right)  =0\ \label{3.1}%
\end{equation}

\noindent To prove Eq.(\ref{3.1}) by recurrence relations, we note that for
the case of $a=0,\ c=\frac{t}{2}+1,\ x=\frac{\tilde{t}}{2}$, Eq.(\ref{A1})
says
\begin{equation}
U\left(  -1,\frac{t}{2}+1,\frac{\tilde{t}}{2}\right)  +U\left(  0,\frac{t}%
{2}+1,\frac{\tilde{t}}{2}\right)  =0\ . \label{3.2}%
\end{equation}
We then apply Eq.(\ref{R3}) for the second term of Eq.(\ref{3.2}) to obtain
Eq.(\ref{3.1}). This completes the proof of Regge stringy Ward identity
Eq.(\ref{W2.1}) based on recurrence relations Eq.(\ref{A1}) and Eq.(\ref{A2}).
The Ward identity in Eq.(\ref{W2.2}) implies
\begin{equation}
\frac{1}{\sqrt{2}}\frac{\tilde{t}^{\prime}}{2}\left[  U\left(  0,\frac{t}%
{2}+2,\frac{\tilde{t}}{2}\right)  +U\left(  -1,\frac{t}{2}+1,\frac{\tilde{t}%
}{2}\right)  \right]  +\left(  -\frac{t}{2}-1\right)  U\left(  -1,\frac{t}%
{2},\frac{\tilde{t}}{2}\right)  =0. \label{3.3}%
\end{equation}

\noindent To prove Eq.(\ref{3.3}) by using recurrence relations, we note that
Eq.(\ref{3.1}) implies the first and the second terms of Eq.(\ref{3.3}) cancel
out. Eq.(\ref{R3}) and $t=\tilde{t}$ say that the last term of Eq.(\ref{3.3})
vanishes. Finally, to prove Eq.(\ref{W2.3}) by using recurrence relations, one
needs to prove%
\begin{align}
&  \left[  \frac{1}{10}\left(  \frac{\tilde{t}^{\prime}}{2}\right)  ^{2}%
+\frac{\tilde{t}}{2}-\frac{t}{5}\right]  U\left(  0,\frac{t}{2}+2,\frac
{\tilde{t}}{2}\right)  +\frac{1}{2}U\left(  -2,\frac{t}{2},\frac{\tilde{t}}%
{2}\right) \nonumber\\
&  \quad+\frac{1}{5}\left(  \frac{\tilde{t}^{\prime}}{2}\right)  \left(
-\frac{t}{2}-1\right)  U\left(  0,\frac{t}{2}+1,\frac{\tilde{t}}{2}\right)
+\frac{1}{10}\left(  -\frac{t}{2}-1\right)  \left(  -\frac{t}{2}\right)
U\left(  0,\frac{t}{2},\frac{\tilde{t}}{2}\right)  =0. \label{3.4}%
\end{align}

\noindent Now Eq.(\ref{R2}) implies
\begin{equation}
U\left(  0,\frac{t}{2}+2,\frac{\tilde{t}}{2}\right)  =U\left(  0,\frac{t}%
{2}+1,\frac{\tilde{t}}{2}\right)  =U\left(  0,\frac{t}{2},\frac{\tilde{t}}%
{2}\right)  . \label{3.5}%
\end{equation}
Therefore Eq.(\ref{3.4}) is equivalent to
\begin{equation}
\frac{t}{2}U\left(  0,\frac{t}{2},\frac{\tilde{t}}{2}\right)  +U\left(
-2,\frac{t}{2},\frac{\tilde{t}}{2}\right)  =0. \label{3.6}%
\end{equation}
Finally one can use Eq.(\ref{R2}) and $\tilde{t}=t$ to prove Eq.(\ref{3.6}).
This completes the proof of Regge stringy Ward identities for mass level
$M^{2}=2$ by using recurrence relations of Kummer functions.

We now turn to the case of mass level $M^{2}=4.$ There are $22$ Regge string
amplitudes for the mass level $M^{2}=4$, $T^{PPP}$,$\ T^{PPL}$, $T^{PPT}$,
$T^{PLL}$, $T^{PLT}$, $T^{PTT}$, $T^{PP}$, $T^{LP}$, $T^{TP}$, $T^{LLL}$,
$T^{LLT}$, $T^{LTT}$, $T^{TTT}$,$T^{PL}$, $T^{LL}$, $T^{TL}$, $T^{PT}$,
$T^{LT}$, $T^{TT}$, $T^{P}$, $T^{L}$, $T^{T}.$ To fix the notation, we adopt
the convention of mass ordered in the $\alpha_{-n}^{\alpha}$ operators, for
example, $T^{LT}(\alpha_{-2}^{L}\alpha_{-1}^{T})$ and $T^{TL}(\alpha_{-2}%
^{T}\alpha_{-1}^{L})$ etc. For this mass level $\tilde{t}=t-2,$ $\tilde
{t}^{\prime}=t+6.$ The $8$ RZNS Eqs.(\ref{R4.1}), (\ref{R4.2}), (\ref{R4.3}),
(\ref{R4.4}), (\ref{R4.5}), (\ref{R4.6}), (\ref{R4.7}) and (\ref{R4.8})
calculated in the appendix B give $8$ Regge stringy Ward identities%
\begin{equation}
25T^{PPP}+9T^{PLL}+9T^{PTT}-9T^{LL}-9T^{TT}-75T^{PP}+50T^{P}=0, \label{R4}%
\end{equation}%
\begin{equation}
T^{PLL}-T^{LL}=0, \label{R5}%
\end{equation}%
\begin{equation}
T^{PTT}-T^{TT}=0, \label{R6}%
\end{equation}%
\begin{equation}
T^{PLT}-T^{(LT)}=0, \label{R7}%
\end{equation}%
\begin{equation}
9T^{PPT}+T^{LLT}+T^{TTT}-18T^{(PT)}+6T^{T}=0, \label{R8}%
\end{equation}%
\begin{equation}
9T^{PPL}+T^{LLL}+T^{LTT}-18T^{(PL)}+6T^{L}=0, \label{R9}%
\end{equation}
\
\begin{equation}
T^{LLT}+T^{TTT}-9T^{[PT]}-3T^{T}=0, \label{R10}%
\end{equation}%
\begin{equation}
T^{LLL}+T^{LTT}-9T^{[PL]}-3T^{L}=0. \label{R11}%
\end{equation}
It is obvious to see that these eight Regge stringy Ward identities are not
good enough to solve the $22$ Regge string scattering amplitudes
algebraically. Indeed, for example, the amplitude $T^{[LT]}$ does not even
show up in any of these eight Ward identities. However, in the GR, one can
identify $e^{P}$ and $e^{L}$ components \cite{ChanLee1,ChanLee2}
(Correspondingly the creation operators $\alpha^{P}_{-n} $ and $-\alpha
^{L}_{-n} $ are identified, where the sign comes from the difference between
the timelike and spacelike directions specified by the metric of the
scattering plane $\eta_{\mu\nu}=diag (-1,1,1)$ .), and take high-energy fixed
angle limit to get three Ward identities in leading order energy
\cite{ChanLee1,ChanLee2}
\begin{align}
T^{LLT}+T^{(LT)}  &  =0,\\
10T^{LLT}+T^{TTT}+18T^{(LT)}  &  =0,\\
T^{LLT}+T^{TTT}+9T^{[LT]}  &  =0,
\end{align}
which can be easily solved to get \cite{ChanLee1,ChanLee2}
\begin{equation}
T^{TTT}:T^{LLT}:T^{(LT)}:T^{[LT]}=8:1:-1:-1.
\end{equation}
The ratios above are consistent with Eq.(\ref{3}).

\bigskip For illustration, we now proceed to prove Regge stringy Ward
identities Eq.(\ref{R4}) to Eq.(\ref{R7}) by using recurrence relations
Eq.(\ref{A1}), Eq.(\ref{A2}) and%
\begin{equation}
\left(  c-a-1\right)  U(a,c-1,x)-\left(  x+c-1\right)  U\left(  a,c,x\right)
+xU\left(  a,c+1,x\right)  =0. \label{A4}%
\end{equation}
Other Regge stringy Ward identities Eq.(\ref{R8}) to Eq.(\ref{R11}) can be
similarly proved by using recurrence relations. For the case of $a=-1$ ,
$c=x+1$, Eq.(\ref{A4}) reduces to%

\begin{equation}
\left(  x+1\right)  U\left(  -1,x,x\right)  -2xU\left(  -1,x+1,x\right)
+xU\left(  -1,x+2,x\right)  =0. \label{R4-1}%
\end{equation}
For the case of $a=-1$, $c=x+2$, Eq.(\ref{A2}) reduces to
\begin{equation}
U\left(  -1,x+2,x\right)  +U\left(  0,x+2,x\right)  -U\left(  -1,x+1,x\right)
=0. \label{R4-2}%
\end{equation}
Finally Eq.(\ref{R4-1}), Eq.(\ref{R4-2}), and Eq.(\ref{R1}) say
\begin{align}
U(-1,x+2,x)  &  =-2U\left(  0,x+2,x\right)  ,\label{R4-3}\\
U\left(  -1,x+1,x\right)   &  =-U\left(  0,x+2,x\right)  . \label{R4-4}%
\end{align}

We are now ready to prove Regge stringy Ward identities. We first prove Regge
stringy Ward identity Eq.(\ref{R5}). The two terms in Eq.(\ref{R5}) divided by
the beta function can be calculated to be
\begin{align}
\frac{1}{B}T^{PLL}  &  =-\frac{1}{M}\left(  \frac{\tilde{t}^{\prime}}%
{2M}\right)  ^{2}\left[  U\left(  -1,\frac{t}{2}+1,\frac{t}{2}-1\right)
+2\left(  \frac{2}{\tilde{t^{\prime}}}\right)  \left(  -\frac{t}{2}-1\right)
U\left(  -1,\frac{t}{2},\frac{t}{2}-1\right)  \right. \nonumber\\
&  \quad\qquad\qquad\qquad\left.  +\left(  \frac{2}{\tilde{t^{\prime}}%
}\right)  ^{2}\left(  -\frac{t}{2}-1\right)  \left(  -\frac{t}{2}\right)
U\left(  -1,\frac{t}{2}-1,\frac{t}{2}-1\right)  \right] \nonumber\\
&  =-\frac{1}{M}\left(  \frac{t+6}{2M}\right)  ^{2}\left[
\begin{array}
[c]{c}%
U\left(  -1,\frac{t}{2}+1,\frac{t}{2}-1\right)  -2\frac{t+2}{t+6}U\left(
-1,\frac{t}{2},\frac{t}{2}-1\right) \\
+\frac{t\left(  t+2\right)  }{\left(  t+6\right)  ^{2}}U\left(  -1,\frac{t}%
{2}-1,\frac{t}{2}-1\right)
\end{array}
\right]  , \label{3.7}%
\end{align}%
\begin{align}
\frac{1}{B}T^{LL}  &  =\left(  \frac{\tilde{t}^{\prime}}{2M}\right)
^{2}\left[  U\left(  0,\frac{t}{2}+2,\frac{t}{2}-1\right)  +\left(  \frac
{2}{\tilde{t^{\prime}}}\right)  \left(  -\frac{t}{2}-1\right)  U\left(
0,\frac{t}{2}+1,\frac{t}{2}-1\right)  \right] \nonumber\\
&  =\left(  \frac{t+6}{2M}\right)  ^{2}\left[  U\left(  0,\frac{t}{2}%
+2,\frac{t}{2}-1\right)  -\frac{t+2}{t+6}U\left(  0,\frac{t}{2}+1,\frac{t}%
{2}-1\right)  \right]  \ . \label{3.8}%
\end{align}
Therefore we want to show
\begin{align}
&  -\frac{1}{M}\left[  U\left(  -1,\frac{t}{2}+1,\frac{t}{2}-1\right)
-2\frac{t+2}{t+6}U\left(  -1,\frac{t}{2},\frac{t}{2}-1\right)  +\frac{t\left(
t+2\right)  }{\left(  t+6\right)  ^{2}}U\left(  -1,\frac{t}{2}-1,\frac{t}%
{2}-1\right)  \right] \nonumber\\
&  \qquad\qquad-U\left(  0,\frac{t}{2}+2,\frac{t}{2}-1\right)  +\frac
{t+2}{t+6}U\left(  0,\frac{t}{2}+1,\frac{t}{2}-1\right)  \overset{?}{=}0
\label{3.9}%
\end{align}
Eq.(\ref{R1}) implies the third term of Eq.(\ref{3.9}) vanishes and therefore
Eq.(\ref{R3}) implies that Eq.(\ref{3.9}) is equivalent to
\begin{align}
&  -\frac{1}{M}U\left(  -1,\frac{t}{2}+1,\frac{t}{2}-1\right)  +\frac{2}%
{M}\frac{t+2}{t+6}U\left(  -1,\frac{t}{2},\frac{t}{2}-1\right)  -\frac{4}%
{t+6}U\left(  0,\frac{t}{2}+1,\frac{t}{2}-1\right) \nonumber\\
&  =\frac{1}{M}\left[  -U\left(  -1,\frac{t}{2}+1,\frac{t}{2}-1\right)
+2\frac{t+2}{t+6}U\left(  -1,\frac{t}{2},\frac{t}{2}-1\right)  -\frac{8}%
{t+6}U\left(  0,\frac{t}{2}+1,\frac{t}{2}-1\right)  \right] \nonumber\\
&  =0 \label{3.10}%
\end{align}
For the case of $x=\frac{t}{2}-1$, Eq.(\ref{R4-3}) and Eq.(\ref{R4-4})
implies
\begin{equation}
U\left(  -1,\frac{t}{2}+1,\frac{t}{2}-1\right)  =2U\left(  -1,\frac{t}%
{2},\frac{t}{2}-1\right)  =-2U\left(  0,\frac{t}{2}+1,\frac{t}{2}-1\right)
\ . \label{3.11}%
\end{equation}
Hence Eq.(\ref{3.10}) is easily proved.

We now prove Regge stringy Ward identity Eq.(\ref{R6}). The two terms in
Eq.(\ref{R6}) divided by the beta function can be calculated to be
\[
\frac{1}{B}T^{PTT}=\left(  -t\right)  \left(  -\frac{1}{M}\right)  U\left(
-1,\frac{t}{2}+1,\frac{t}{2}-1\right)  ,
\]%
\[
\frac{1}{B}T^{TT}=\left(  -t\right)  U\left(  0,\frac{t}{2}+2,\frac{t}%
{2}-1\right)  .
\]
Therefore we want to show
\begin{equation}
\frac{t}{M}U\left(  -1,\frac{t}{2}+1,\frac{t}{2}-1\right)  +tU\left(
0,\frac{t}{2}+2,\frac{t}{2}-1\right)  \overset{?}{=}0 \label{3.12}%
\end{equation}
For the case of $x=\frac{t}{2}-1$, Eq.(\ref{R4-3}) means
\begin{equation}
U\left(  -1,\frac{t}{2}+1,\frac{t}{2}-1\right)  =-2U\left(  0,\frac{t}%
{2}+1,\frac{t}{2}-1\right)  \ . \label{3.13}%
\end{equation}
Eq.(\ref{3.13}) and Eq.(\ref{R3}) prove Eq.(\ref{3.12}).

We can now turn to prove Regge stringy Ward identity Eq.(\ref{R4}). We first
note that Eq.(\ref{R5}) and Eq.(\ref{R6}) implies that Eq.(\ref{R4}) is
equivalent to
\begin{equation}
25T^{PPP}-75T^{PP}+50T^{P}=0. \label{3.14}%
\end{equation}
The three terms in Eq.(\ref{3.14}) divided by the beta function are
\begin{align}
\frac{1}{B}T^{P}  &  =\left(  -\frac{t-2}{2M}\right)  U\left(  0,\frac{t}%
{2}+2,\frac{t}{2}-1\right)  ,\label{3.15}\\
\frac{1}{B}T^{PP}  &  =-\frac{1}{M}\left(  -\frac{t-2}{2M}\right)  U\left(
-1,\frac{t}{2}+1,\frac{t}{2}-1\right)  ,\label{3.16}\\
\frac{1}{B}T^{PPP}  &  =\left(  -\frac{1}{M}\right)  ^{3}U\left(  -3,\frac
{t}{2}-1,\frac{t}{2}-1\right)  \ . \label{3.17}%
\end{align}
Therefore we want to show
\begin{align}
&  2\left(  -\frac{t-2}{2M}\right)  U\left(  0,\frac{t}{2}+2,\frac{t}%
{2}-1\right)  -3\left(  -\frac{1}{M}\right)  \left(  -\frac{t-2}{2M}\right)
U\left(  -1,\frac{t}{2}+1,\frac{t}{2}-1\right) \nonumber\\
&  +\left(  -\frac{1}{M}\right)  ^{3}U\left(  -3,\frac{t}{2}-1,\frac{t}%
{2}-1\right)  \overset{?}{=}0 \label{3.18}%
\end{align}
For the case of $a=-2$, $c=x$, Eq.(\ref{A1}) gives%
\begin{equation}
U\left(  -3,x,x\right)  +4U\left(  -2,x,x\right)  +2\left(  1+x\right)
U\left(  -1,x,x\right)  =0. \label{3.19}%
\end{equation}
Using Eq.(\ref{R1}), we obtain
\begin{equation}
U\left(  -3,x,x\right)  +4U\left(  -2,x,x\right)  =0\ . \label{3.20}%
\end{equation}
From Eq.(\ref{3.20}) and Eq.(\ref{R2}), we obtain
\begin{equation}
U\left(  -3,x,x\right)  -4xU\left(  0,x,x\right)  =0\ . \label{3.21}%
\end{equation}
From Eq.(\ref{3.21}) and Eq.(\ref{R3}), we obtain%
\begin{align}
&  2\left(  -\frac{t-2}{2M}\right)  U\left(  0,\frac{t}{2}+2,\frac{t}%
{2}-1\right)  -3\left(  -\frac{1}{M}\right)  \left(  -\frac{t-2}{2M}\right)
U\left(  -1,\frac{t}{2}+1,\frac{t}{2}-1\right) \nonumber\\
&  +\left(  -\frac{1}{M}\right)  ^{3}U\left(  -3,\frac{t}{2}-1,\frac{t}%
{2}-1\right) \nonumber\\
=  &  \left(  2\left(  -\frac{t-2}{2M}\right)  +4\left(  \frac{t-2}{2}\right)
\left(  -\frac{1}{M}\right)  ^{3}\right)  U\left(  0,\frac{t}{2}+2,\frac{t}%
{2}-1\right) \nonumber\\
&  +3\frac{1}{M}\left(  -\frac{t-2}{2M}\right)  U\left(  -1,\frac{t}%
{2}+1,\frac{t}{2}-1\right) \nonumber\\
=  &  \left(  2\left(  -\frac{t-2}{4}\right)  +\left(  \frac{t-2}{2}\right)
\left(  -\frac{1}{2}\right)  \right)  U\left(  0,\frac{t}{2}+2,\frac{t}%
{2}-1\right) \nonumber\\
&  +\frac{3}{2}\left(  -\frac{t-2}{4}\right)  U\left(  -1,\frac{t}{2}%
+1,\frac{t}{2}-1\right) \nonumber\\
=  &  \frac{t-2}{2}\left[  -\frac{3}{2}U\left(  0,\frac{t}{2}+2,\frac{t}%
{2}-1\right)  -\frac{3}{2}\frac{1}{2}U\left(  -1,\frac{t}{2}+1,\frac{t}%
{2}-1\right)  \right]  \ . \label{3.22}%
\end{align}
Finally Eq.(\ref{R3}) and Eq.(\ref{R4-3}) implies that Eq.(\ref{3.22})
vanishes. This proves Eq.(\ref{3.18}).

For the fourth stringy Ward identity at mass level $M^{2}=4$, the two terms in
Eq.(\ref{R7}) divided by the beta function are
\begin{align}
\frac{1}{B}T^{PLT}=  &  -\frac{\sqrt{-t}\tilde{t}^{\prime}}{2M^{2}}\left[
U\left(  -1,\frac{t}{2}+1,\frac{\tilde{t}}{2}\right)  +\left(  \frac{2}%
{\tilde{t}^{\prime}}\right)  \left(  -\frac{t}{2}-1\right)  U\left(
-1,\frac{t}{2},\frac{\tilde{t}}{2}\right)  \right] \nonumber\\
=  &  -\frac{\sqrt{-t}\left(  t+6\right)  }{2M^{2}}\left[
\begin{array}
[c]{c}%
U\left(  -1,\frac{t}{2}+1,\frac{t}{2}-1\right) \\
+\left(  \frac{2}{t+6}\right)  \left(  -\frac{t}{2}-1\right)  U\left(
-1,\frac{t}{2},\frac{t}{2}-1\right)
\end{array}
\right]  ,\label{3.23}\\
\frac{1}{B}T^{LT}=  &  \sqrt{-t}\frac{\tilde{t}^{\prime}}{2M}U\left(
0,\frac{t}{2}+2,\frac{\tilde{t}}{2}\right)  =\sqrt{-t}\frac{t+6}{2M}U\left(
0,\frac{t}{2}+2,\frac{t}{2}-1\right)  ,\label{3.24}\\
\frac{1}{B}T^{TL}=  &  \sqrt{-t}\frac{\tilde{t}^{\prime}}{2M}\left[  U\left(
0,\frac{t}{2}+2,\frac{\tilde{t}}{2}\right)  +\left(  \frac{2}{\tilde
{t}^{\prime}}\right)  \left(  -\frac{t}{2}-1\right)  U\left(  0,\frac{t}%
{2}+1,\frac{\tilde{t}}{2}\right)  \right] \nonumber\\
=  &  \sqrt{-t}\frac{t+6}{2M}\left[  U\left(  0,\frac{t}{2}+2,\frac{t}%
{2}-1\right)  +\left(  \frac{2}{t+6}\right)  \left(  -\frac{t}{2}-1\right)
U\left(  0,\frac{t}{2}+1,\frac{t}{2}-1\right)  \right]  \ . \label{3.25}%
\end{align}
Therefore we want to show%
\begin{align}
&  2T^{PLT}-T^{LT}-T^{TL}\nonumber\\
&  =B\sqrt{t}\left[  -\frac{t+6}{M^{2}}U\left(  -1,\frac{t}{2}+1,\frac{t}%
{2}-1\right)  -\frac{2}{M^{2}}\left(  -\frac{t}{2}-1\right)  U\left(
-1,\frac{t}{2},\frac{t}{2}-1\right)  \right. \nonumber\\
&  \quad\left.  -\frac{t+6}{M}U\left(  0,\frac{t}{2}+2,\frac{t}{2}-1\right)
-\frac{1}{M}\left(  -\frac{t}{2}-1\right)  U\left(  0,\frac{t}{2}+1,\frac
{t}{2}-1\right)  \right]  \overset{?}{=}0\ . \label{3.26}%
\end{align}
Using Eq.(\ref{R3}), we obtain
\begin{align}
&  \frac{1}{B}(2T^{PLT}-T^{LT}-T^{TL})\nonumber\\
&  =\sqrt{t}\left[  -\frac{t+6}{M^{2}}U\left(  -1,\frac{t}{2}+1,\frac{t}%
{2}-1\right)  -\frac{2}{M^{2}}\left(  -\frac{t}{2}-1\right)  U\left(
-1,\frac{t}{2},\frac{t}{2}-1\right)  \right. \nonumber\\
&  \qquad\left.  +\frac{1}{M}\left(  -t-6+\frac{t}{2}+1\right)  U\left(
0,\frac{t}{2},\frac{t}{2}-1\right)  \right] \nonumber\\
&  =\sqrt{t}\left[
\begin{array}
[c]{c}%
-\frac{t+6}{M^{2}}U\left(  -1,\frac{t}{2}+1,\frac{t}{2}-1\right)  -\frac
{2}{M^{2}}\left(  -\frac{t}{2}-1\right)  U\left(  -1,\frac{t}{2},\frac{t}%
{2}-1\right) \\
+\frac{1}{M}\left(  -\frac{t}{2}-5\right)  U\left(  0,\frac{t}{2},\frac{t}%
{2}-1\right)
\end{array}
\right] \nonumber\\
&  =\sqrt{t}\left[
\begin{array}
[c]{c}%
-\frac{t+6}{4}U\left(  -1,\frac{t}{2}+1,\frac{t}{2}-1\right)  +\frac{t+2}%
{4}U\left(  -1,\frac{t}{2},\frac{t}{2}-1\right) \\
-\frac{t+10}{4}U\left(  0,\frac{t}{2},\frac{t}{2}-1\right)
\end{array}
\right]  \label{3.27}%
\end{align}
One can now use Eq.(\ref{R4-3}) and Eq.(\ref{R4-4}) to prove that
Eq.(\ref{3.27}) vanishes. This completes the explicit proof of four Regge
stringy Ward identities for mass level $M^{2}=4$ by using recurrence relations
of Kummer functions. Other four Regge stringy Ward identities can be similarly proved.

We observe that the recurrence relations of Kummer functions are more powerful
than Regge stringy Ward identities in relating Regge string scattering
amplitudes. This is indeed the case as we will show now in the following that
all Regge string scattering amplitudes can be algebraically solved by using
recurrence relations up to multiplicative factors in the first line of
Eq.(\ref{factor1}) (Eq.(\ref{factor2})).

To be more precise, we will first show that the ratio%
\begin{equation}
\frac{U(a,c,x)}{U(0,x,x)}=f(a,c,x),a=0,-1,-2,-3,... \label{Lemma}%
\end{equation}
is fixed and can be calculated by using recurrence relations Eq.(\ref{A1}),
Eq.(\ref{A2}) and%
\begin{equation}
(c-a)U(a,c,x)+U(a-1,c,x)-xU(a,c+1,x)=0. \label{A3}%
\end{equation}
We stress that Eq.(\ref{Lemma}) is nontrivial in the sense that, for general
values of $a$, the best one can obtain from recurrence relations is to express
any Kummer function in terms of any two of its associated function (see
Appendix A). However, Eq.(\ref{Lemma}) states that for nonpositive integer
values of $a,$ $U(a,c,x)$ can be fixed up to an overall factor by using
recurrence relations.

To prove Eq.(\ref{Lemma}), we first note that, for $a=0,c=x,$ recurrence
relation Eq.(\ref{A1}) implies Eq.(\ref{R1}). \ This determines $\frac
{U(a,x,x)}{U(0,x,x)}$ for $a$ is a nonpositive integer$.$ For illustration, we
list examples of relations%
\begin{align*}
a  &  =-1,U(-2,x,x)+0+xU(0,x,x)=0,\\
a  &  =-2,U(-3,x,x)+4U(-2,x,x)+0=0,\\
a  &  =-3,U(-4,x,x)+6U(-3,x,x)+3(2+x)U(-2,x,x)=0,\\
&  ...............
\end{align*}
which determines $\frac{U(-2,x,x)}{U(0,x,x)},\frac{U(-3,x,x)}{U(0,x,x)}%
,\frac{U(-4,x,x)}{U(0,x,x)},....$ recursively.

Next we extend the result to $\frac{U(a,c,x)}{U(0,x,x)}$ for $c=x+Z,Z=$
integer. We first consider the simple case with $a=0$. From Eq.(\ref{A2}), we
obtain for $a=0,c=x+i,i\in Z$%
\begin{equation}
U(0,x+i,x)-U(0,x+i-1,x)=0,
\end{equation}
which gives $\frac{U(0,x+i,x)}{U(0,x,x)}=1.$ This proves Eq.(\ref{Lemma}) for
$a=0.$ For $a\in Z_{-},c=x+Z_{-}$, we obtain from Eq.(\ref{A2}) with $c=x-i$%
\begin{equation}
U(a,x-i,x)-aU(a+1,x-i,x)-U(a,x-i-1,x)=0.
\end{equation}
Since $\frac{U(a,x,x)}{U(0,x,x)},\frac{U(a+1,x,x)}{U(0,x,x)}$ have been
determined for $a\in Z_{-},$ this determines $\frac{U(a,x-i,x)}{U(0,x,x)}$ for
$a\in Z_{-},i=1,2,3...$ recursively. For $a\in Z_{-},c=x+Z_{+},$ we obtain
from Eq.(\ref{A3}) with $c=x+i$%
\begin{equation}
(x-a+i)U(a,x+i,x)+U(a-1,x+i,x)-xU(a,x+i+1,x)=0.
\end{equation}
Since $\frac{U(a-1,x,x)}{U(0,x,x)},\frac{U(a,x,x)}{U(0,x,x)}$ have been
determined for $a\in Z_{-},$ this determines $\frac{U(a,x+i,x)}{U(0,x,x)}$ for
$a\in Z_{-},i=1,2,3...$ recursively. This completes the proof of
Eq.(\ref{Lemma}) by using recurrence relations of Kummer functions.

Secondly, we want to show that each Kummer function in the summation of
Eq.(\ref{factor2}) can be expressed in terms of Regge string scattering
amplitudes. To show this, we first consider $r_{1}=0$ amplitudes in a fixed
mass level and a fixed $q_{1}$ with no summation over Kummer functions. These
amplitudes contain only one Kummer function. Then let us take the amplitude
with the maximum $p_{1}$. By decreasing $p_{1}$ and increasing $r_{1}$ by $1$,
we can create an amplitude with two Kummer functions in the same mass level
and the same $q_{1}$. The first one of the two Kummer functions is the one
appeared in the previous amplitude with $r_{1}=0$, so we can write the second
Kummer function in terms of the two amplitudes, one with $r_{1}=0$ and the
other with $r_{1}=1$. By decreasing $p_{1}$ and increasing $r_{1}$ by $1$
again, we can create an amplitude with three Kummer functions in the same mass
level and the same $q_{1}$. The first two of the three Kummer functions is the
ones appeared in the previous two amplitudes, so we can write the third Kummer
functions in terms of the three amplitudes. We can repeat this process until
$p_{1}=0$. In this way, we can express all the Kummer functions in
Eq.(\ref{factor2}) in terms of the RR amplitudes. 

In the following, as an example, let us illustrate the above process for the
mass level $4$ amplitudes. There are $22$ Regge string amplitudes for the mass
level $M^{2}=4.$ We first consider the group of amplitudes with $q_{1}=0,$
$(T^{TTT},T^{LTT},T^{LLT},T^{LLL})$. The corresponding $r_{1}$ for each
amplitude are $(0,1,2,3)$. By using Eq.(\ref{factor2}), one can easily see
that $U\left(  0,\frac{t}{2}+2,\frac{t}{2}-1\right)  $ can be expressed in
terms of $T^{TTT}$, $U\left(  0,\frac{t}{2}+1,\frac{t}{2}-1\right)  $ can be
expressed in terms of $(T^{TTT},T^{LTT})$, $U\left(  0,\frac{t}{2},\frac{t}%
{2}-1\right)  $ can be expressed in terms of $(T^{TTT},T^{LTT},T^{LLT})$, and
finally $U\left(  0,\frac{t}{2}-1,\frac{t}{2}-1\right)  $ can be expressed in
terms of $(T^{TTT},T^{LTT},T^{LLT},T^{LLL})$. Similarly, we can consider
groups of amplitudes $(T^{PT},T^{PL})$, $(T^{LT},T^{LL})$ and $(T^{TT}%
,T^{TL})$ with $q_{1}=0$; group of amplitude $(T^{PTT},T^{PLT},T^{PLL})$ with
$q_{1}=1$ and group of amplitude $(T^{PPT},T^{PPL})$ with $q_{1}=2$. All the
remaining $7$ amplitudes are with $r_{1}=0$, and each amplitude contains only
one Kummer function. Due to the multiplicative factors, there are much more RR
amplitudes than the number of Kummer functions involved at each fixed mass
level. At mass level $4$, for example, there are $22$ RR amplitudes and only
$10$ Kummer functions involved. So there is an onto correspondence between RR
amplitudes and Kummer functions. We have done the analysis by using
Eq.(\ref{factor2}). Similar analysis can be performed by using
Eq.(\ref{factor1}) to get the same results.

An important application of the above prescription is the construction of an
infinite number of recurrence relations among Regge string scattering
amplitudes. One can use the recurrence relations of Kummer functions
Eq.(\ref{RC1}) to Eq.(\ref{RC6}) to systematically construct recurrence
relations among Regge string scattering amplitudes.

Note that a simple calculation by using the explicit form of Kummer function
in Eq.(\ref{finite}) gives $U(0,x,x)=1.$ However, when applying to the case of
Regge string scattering amplitudes, it will bring back a multiplicative factor
in the first line of Eq.(\ref{factor1}), (Eq.(\ref{factor2})) for each
amplitude. We thus conclude that all Regge string scattering amplitudes can be
algebraically solved by recurrence relations of Kummer functions up to
multiplicative factors.

Finally we calculate some examples of recurrence relations among Regge string
scattering amplitudes. At mass level $M^{2}=2,$ by using Eq.(\ref{factor2})
and the recurrence relation
\begin{equation}
U\left(  -2,\frac{t}{2},\frac{t}{2}\right)  +\left(  \frac{t}{2}+1\right)
U(-1,\frac{t}{2},\frac{t}{2})-\frac{t}{2}U\left(  -1,\frac{t}{2}+1,\frac{t}%
{2}\right)  =0,
\end{equation}
one can obtain the following recurrence relation among Regge string scattering
amplitudes%
\begin{equation}
M\sqrt{-t}T^{PP}-\frac{t}{2}T^{PT}=0. \label{RRT1}%
\end{equation}
In contrast to the Regge stringy Ward identities Eq.(\ref{W2.1}),
Eq.(\ref{W2.2}) and Eq.(\ref{W2.3}) which contain only constant coefficients,
the recurrence relation in Eq.(\ref{RRT1}) contains kinematic variable $t$ in
its coefficients. Note that Eq.(\ref{RRT1}) is independent of all three Regge
stringy Ward identities at mass level $M^{2}=2$.

At mass level $M^{2}=4,$ by using Eq.(\ref{factor2}), one can calculate%
\begin{align}
\frac{1}{B}T^{PPP}=  &  \left(  -\frac{1}{M}\right)  ^{3}U\left(  -3,\frac
{t}{2}-1,\frac{t}{2}-1\right)  ,\\
\frac{1}{B}T^{PPT}=  &  \left(  -\frac{1}{M}\right)  ^{2}\sqrt{-t}U\left(
-2,\frac{t}{2},\frac{t}{2}-1\right)  ,\\
\frac{1}{B}T^{PPL}=  &  \frac{t+6}{2M^{3}}U\left(  -2,\frac{t}{2},\frac{t}%
{2}-1\right)  +\frac{1}{M^{3}}\left(  -\frac{t}{2}-1\right)  U\left(
-2,\frac{t}{2}-1,\frac{t}{2}-1\right)  .
\end{align}
The recurrence relation%
\begin{equation}
U\left(  -3,\frac{t}{2}-1,\frac{t}{2}-1\right)  +\left(  \frac{t}{2}+1\right)
U(-2,\frac{t}{2}-1,\frac{t}{2}-1)-(\frac{t}{2}-1)U\left(  -2,\frac{t}{2}%
,\frac{t}{2}-1\right)  =0 \label{RRTT}%
\end{equation}
leads to the following recurrence relation among Regge string scattering
amplitudes%
\begin{equation}
M\sqrt{-t}T^{PPP}-4T^{PPT}+M\sqrt{-t}T^{PPL}=0. \label{RRT2}%
\end{equation}
We have explicitly verified Eq.(\ref{RRT1}) and Eq.(\ref{RRT2}). It will be
difficult to identify identity like Eq.(\ref{RRT2}) without using the
recurrence relation Eq.(\ref{RRTT}). One can similarly construct infinite
number of them for amplitudes at arbitrary higher mass levels based on the
recurrence relations of Kummer functions and their associated functions (see
Appendix A).

The recurrence relations Eq.(\ref{RRT1}) and Eq.(\ref{RRT2}) for Regge string
scattering amplitudes are reminiscent of four point BCJ relations
\cite{BCJ1,BCJ2,BCJ3,BCJ4,BCJ5} for Yang-Mills gluon color-stripped scattering
amplitudes $A$%
\begin{align}
tA(k_{1},k_{4},k_{2},k_{3})-sA(k_{1},k_{3},k_{4},k_{2})  &  =0,\nonumber\\
sA(k_{1},k_{2},k_{3},k_{4})-uA(k_{1},k_{4},k_{2},k_{3})  &  =0,\nonumber\\
\text{ }uA(k_{1},k_{3},k_{4},k_{2})-tA(k_{1},k_{2},k_{3},k_{4})  &  =0
\label{BCJ}%
\end{align}
where non-constant coefficients, or kinematic variables, show up in the
relations. Note that other relation such as four point KK relation
\cite{KK1,KK2}
\begin{equation}
A(k_{1},k_{2},k_{3},k_{4})+A(k_{1},k_{3},k_{4},k_{2})+A(k_{1},k_{4}%
,k_{2},k_{3})=0 \label{KK}%
\end{equation}
contains only constant coefficients which is similar to Regge stringy Ward
identities.\qquad

\bigskip For the third example, we construct an inter-mass level recurrence
relation for Regge string scattering amplitudes at mass level $M^{2}=2,4.$ We
begin with the addition theorem of Kummer function \cite{Slater}%
\begin{equation}
U(a,c,x+y)=\sum_{k=0}^{\infty}\frac{1}{k!}\left(  a\right)  _{k}(-1)^{k}%
y^{k}U(a+k,c+k,x)
\end{equation}
which terminates to a finite sum for a nonpositive integer $a.$ By taking, for
example, $a=-1,c=\frac{t}{2}+1,x=\frac{t}{2}-1$ and $y=1,$ the theorem gives%
\begin{equation}
U\left(  -1,\frac{t}{2}+1,\frac{t}{2}\right)  -U(-1,\frac{t}{2}+1,\frac{t}%
{2}-1)-U\left(  0,\frac{t}{2}+2,\frac{t}{2}-1\right)  =0. \label{inter}%
\end{equation}
Note that, unlike all previous cases, the last arguments of Kummer functions
in Eq.(\ref{inter}) can be different. Eq.(\ref{inter}) leads to an inter-mass
level recurrence relation%
\begin{equation}
M(2)(t+6)T_{2}^{TP}-2M(4)^{2}\sqrt{-t}T_{4}^{LP}+2M(4)T_{4}^{LT}=0
\label{RRT3}%
\end{equation}
where \ masses $M(2)=\sqrt{2},M(4)=\sqrt{4}=2,$ and $T_{2},T_{4}$ are Regge
string scattering amplitudes for mass levels $M^{2}=2,4$ respectively. In
deriving Eq.(\ref{RRT3}), it is important to use the fact that the Regge power
law behavior in Eq.(\ref{power}) is universal and is mass level independent
\cite{bosonic}.

Following the same procedure, one can construct infinite number of recurrence
relations among Regge string scattering amplitudes at arbitrary mass levels
which, in general, are independent of Regge stringy Ward identities. The
physical origin for the four point BCJ relations Eq.(\ref{BCJ}), for example,
can be traced back to the conservation of momenta. On the contrary, the
physical origin of these new recurrence relations among Regge string
scattering amplitudes or "symmetries" is not well understood and is an
interesting problem to study.

\section{Conclusion}

In this paper, we calculate the complete set of high-energy string scattering
amplitudes in the Regge regime. We derive Regge stringy Ward identities for
the first few mass levels based on the decoupling of zero-norm states. These
results are valid even for higher point functions and higher point loops by
unitarity. We found that, unlike the case for the fixed angle regime, the
Regge stringy Ward identities were not good enough to solve all the Regge
string scattering amplitudes algebraically. On the other hand, we found that
all the Regge stringy Ward identities can be explicitly proved by the
recurrence relations of Kummer functions of the second kind. We then show
that, instead of Regge stringy Ward identities, one can use these recurrence
relations to solve all Regge string scattering amplitudes algebraically up to
multiplicative factors.

Finally, for illustration, we calculate some examples of recurrence relations
among Regge string scattering amplitudes of different string states based on
recurrence relations and addition theorem of Kummer functions. In contrast to
the Regge stringy Ward identities which contain only constant coefficients,
these recurrence relations contains kinematic variable $t$ in its coefficients
and are in general independent of Regge stringy Ward identities. The dynamical
origin of these recurrence relations remain to be studied. These recurrence
relations among Regge string scattering amplitudes are dual to linear
relations or symmetries among high-energy fixed angle string scattering
amplitudes discovered previously \cite{ChanLee1,ChanLee2,PRL,CHLTY}.

Recently, five-point tachyon amplitude was considered in the context of BCFW
application of string theory in \cite{BCFW}. It will be interesting to
consider both RR and GR of higher spin five-point scattering amplitudes.

\section{Acknowledgments}

We thank Yung-Yeh Chang, Chih-Hao Fu, Song He, Yu-Ting Huang, Chung-I Tan and
Yi Yang for helpful discussions. This work is supported in part by the
National Science Council, 50 billions project of Ministry of Education,
National Center for Theoretical Sciences and S.T. Yau center of NCTU, Taiwan.

\appendix%

\setcounter{equation}{0}
\renewcommand{\theequation}{\thesection.\arabic{equation}}%

\section{Recurrence Relations of Kummer Functions}

In this appendix, we review the recurrence relations of Kummer functions of
the second kind \cite{Slater}. The Kummer function of the second kind $U$ \ is
defined to be%
\begin{equation}
U(a,c,x)=\frac{\pi}{\sin\pi c}\left[  \frac{M(a,c,x)}{(a-c)!(c-1)!}%
-\frac{x^{1-c}M(a+1-c,2-c,x)}{(a-1)!(1-c)!}\right]  \text{ \ }(c\neq2,3,4...)
\end{equation}
where $M(a,c,x)=\sum_{j=0}^{\infty}\frac{(a)_{j}}{(c)_{j}}\frac{x^{j}}{j!}$ is
the Kummer function of the first kind. Here $(a)_{j}=a(a+1)(a+2)...(a+j-1)$ is
the Pochhammer symbol. $U$ and $M$ are the two solutions of the Kummer
Equation%
\begin{equation}
xy^{^{\prime\prime}}(x)+(c-x)y^{\prime}(x)-ay(x)=0. \label{KE}%
\end{equation}
For any confluent hypergeometric function with parameters $(a,c)$ the four
functions with parameters $(a-1,c),(a+1,c),(a,c-1)$ and $(a,c+1)$ are called
the contiguous functions. It follows, from the Kummer Equation Eq.(\ref{KE})
and derivatives of Kummer functions%
\begin{align}
U(a+1,c+1,x)  &  =\frac{-1}{a}U^{\prime}(a,c,x),\label{DE1}\\
U(a+1,c,x)  &  =\frac{1}{1+a-c}U(a,c,x)+\frac{x}{a(1+a-c)}U^{\prime}(a,c,x),\\
U(a,c-1,x)  &  =\frac{1-c}{1+a-c}U(a,c,x)-\frac{x}{1+a-c}U^{\prime}(a,c,x),\\
U(a,c+1,x)  &  =U^{\prime}(a,c,x)-U^{\prime}(a,c,x),\\
U(a-1,c,x)  &  =(x+a-c)U(a,c,x)-xU^{\prime}(a,c,x),\\
U(a-1,c-1,x)  &  =(1+x-c)U(a,c,x)-xU^{\prime}(a,c,x), \label{DE6}%
\end{align}
that a recurrence relation exists between any such function and any two of its
contiguous functions. There are six recurrence relations%
\begin{align}
U(a-1,c,x)-(2a-c+x)U(a,c,x)+a(1+a-c)U(a+1,c,x)  &  =0,\label{RC1}\\
(c-a-1)U(a,c-1,x)-(x+c-1))U(a,c,x)+xU(a,c+1,x)  &  =0,\label{RC2}\\
U(a,c,x)-aU(a+1,c,x)-U(a,c-1,x)  &  =0,\label{RC3}\\
(c-a)U(a,c,x)+U(a-1,c,x)-xU(a,c+1,x)  &  =0,\label{RC4}\\
(a+x)U(a,c,x)-xU(a,c+1,x)+a(c-a-1)U(a+1,c,x)  &  =0,\label{RC5}\\
(a+x-1)U(a,c,x)-U(a-1,c,x)+(1+a-c)U(a,c-1,x)  &  =0. \label{RC6}%
\end{align}
From any two of these six relations the remaining four recurrence relations
can be deduced. Thus they are not independent. For example, one can deduces
recurrence relation Eq.(\ref{RC1}) from Eq.(\ref{RC3}) and Eq.(\ref{RC4}). We
start with Eq.(\ref{RC3}) with $c\rightarrow c+1$
\begin{equation}
\quad U\left(  a,c+1,x\right)  -aU\left(  a+1,c+1,x\right)  -U(a,c,x)=0\ .
\label{RC7}%
\end{equation}
We consider Eq.(\ref{RC4})+$x\cdot$Eq.(\ref{RC7}) to deduce
\begin{equation}
\quad\left(  c-a-x\right)  U\left(  a,c,x\right)  +U\left(  a-1,c,x\right)
-axU\left(  a+1,c+1,x\right)  =0\ . \label{RC8}%
\end{equation}
Next we replace Eq.(\ref{RC4}) with $a\rightarrow a+1$ to get
\begin{equation}
\quad\left(  c-a-1\right)  U\left(  a+1,c,x\right)  +U\left(  a,c,x\right)
-xU(a+1,c+1,x)=0\ . \label{RC9}%
\end{equation}
Finally we consider Eq.(\ref{RC8})$-a\cdot$Eq.(\ref{RC9}) to deduce
\begin{equation}
\quad\left(  c-2a-x\right)  U\left(  a,c,x\right)  +U\left(  a-1,c,x\right)
-a\left(  c-a-1\right)  U\left(  a+1,c,x\right)  =0\ , \label{RC10}%
\end{equation}
which is nothing but Eq.(\ref{RC1}).

The confluent hypergeometric function with parameters $(a\pm m,c\pm n)$ for
$m,n=0,1,2...$are called associated functions. Again it can be shown that
there exist relations between any three associated functions, so that any
confluent hypergeometric function can be expressed in terms of any two of its
associated functions.

\section{Regge String Zero-Norm States}

There are two types of zero-norm states (ZNS) in the old covariant first
quantized string spectrum%

\begin{equation}
\text{Type I}:L_{-1}\left\vert x\right\rangle ,\text{ where }L_{1}\left\vert
x\right\rangle =L_{2}\left\vert x\right\rangle =0,\text{ }L_{0}\left\vert
x\right\rangle =0; \label{ZN1}%
\end{equation}

\begin{equation}
\text{Type II}:(L_{-2}+\frac{3}{2}L_{-1}^{2})\left\vert \widetilde{x}%
\right\rangle ,\text{ where }L_{1}\left\vert \widetilde{x}\right\rangle
=L_{2}\left\vert \widetilde{x}\right\rangle =0,\text{ }(L_{0}+1)\left\vert
\widetilde{x}\right\rangle =0. \label{ZN2}%
\end{equation}
Eq.(\ref{ZN1}) and Eq.(\ref{ZN2}) can be derived from Kac determinant in
conformal field theory. While type I states have zero-norm at any spacetime
dimension, type II states have zero-norm \textit{only} at D=26. The existence
of type II zero-norm states signals the importance of zero-norm states in the
structure of the theory of string. In fact, by requiring the decoupling of
these two types of zero-norm states or stringy Ward identities in the
high-energy fixed angle regime, one can calculate algebraically the complete
ratios among high-energy string scattering amplitudes of different string
states at each fixed mass level in Eq.(\ref{3}).

In the RR, however, the Regge stringy Ward identities turn out to be not good
enough to solve all the Regge scattering amplitudes algebraically. This is due
to the much more numerous Regge string scattering amplitudes than those in the
GR at each fixed mass level. In this appendix, we list all ZNS for $M^{2}=2$
and $\ 4$ and calculate their Regge limit which we use in the text to
demonstrate the calculation. At the first massive level $k^{2}=-2,$ there is a
type II ZNS%

\begin{equation}
\lbrack\frac{1}{2}\alpha_{-1}\cdot\alpha_{-1}+\frac{5}{2}k\cdot\alpha
_{-2}+\frac{3}{2}(k\cdot\alpha_{-1})^{2}]\left\vert 0,k\right\rangle
\label{2.1}%
\end{equation}
and a type I ZNS%

\begin{equation}
\lbrack\theta\cdot\alpha_{-2}+(k\cdot\alpha_{-1})(\theta\cdot\alpha
_{-1})]\left\vert 0,k\right\rangle ,\theta\cdot k=0. \label{2.2}%
\end{equation}
In the Regge limit, the polarizations of the 2nd particle with momentum
$k_{2}$ on the scattering plane used in the text were defined to be
$e^{P}=\frac{1}{M_{2}}(E_{2},\mathrm{k}_{2},0)=\frac{k_{2}}{M_{2}}$ as the
momentum polarization, $e^{L}=\frac{1}{M_{2}}(\mathrm{k}_{2},E_{2},0)$ the
longitudinal polarization and $e^{T}=(0,0,1)$ the transverse polarization
which lies on the scattering plane. $\eta_{\mu\nu}=diag(-1,1,1).$ The three
vectors $e^{P}$, $e^{L}$ and $e^{T}$ satisfy the completeness relation
$\eta_{\mu\nu}=\sum_{\alpha,\beta}e_{\mu}^{\alpha}e_{\nu}^{\beta}\eta
_{\alpha\beta}$ where $\mu,\nu=0,1,2$ and $\alpha,\beta=P,L,T$ and
$\alpha_{-1}^{T}=\sum_{\mu}e_{\mu}^{T}\alpha_{-1}^{\mu}$, $\alpha_{-1}%
^{T}\alpha_{-2}^{L}=\sum_{\mu,\nu}e_{\mu}^{T}e_{\nu}^{L}\alpha_{-1}^{\mu
}\alpha_{-2}^{\nu}$ etc.

In the Regge limit, the type II ZNS in Eq.(\ref{2.1}) gives the Regge
zero-norm state (RZNS)%
\begin{equation}
(\sqrt{2}\alpha_{-2}^{P}-\alpha_{-1}^{P}\alpha_{-1}^{P}-\frac{1}{5}\alpha
_{-1}^{L}\alpha_{-1}^{L}-\frac{1}{5}\alpha_{-1}^{T}\alpha_{-1}^{T}%
)|0,k\rangle.\label{R2.3}%
\end{equation}
Type I ZNS in Eq.(\ref{2.2}) gives two RZNS%
\begin{equation}
(\alpha_{-2}^{T}-\sqrt{2}\alpha_{-1}^{P}\alpha_{-1}^{T})|0,k\rangle
,\label{R2.1}%
\end{equation}%
\begin{equation}
(\alpha_{-2}^{L}-\sqrt{2}\alpha_{-1}^{P}\alpha_{-1}^{L})|0,k\rangle
.\label{R2.2}%
\end{equation}
RZNS in Eq.(\ref{R2.1}) and Eq.(\ref{R2.2}) correspond to choose $\theta^{\mu
}=e^{T}$ and $\theta^{\mu}=e^{L}$ respectively. Note that the norms of Regge
"zero-norm" states may not be zero. For instance the norm of Eq.(\ref{R2.3})
is not zero. They are just used to produce Regge stringy Ward identities
Eq.(\ref{W2.3}), Eq.(\ref{W2.1}) and Eq.(\ref{W2.2}) in the text.

At the second massive level $k^{2}=-4,$ there is a type I scalar ZNS%
\begin{align}
&  [\frac{17}{4}(k\cdot\alpha_{-1})^{3}+\frac{9}{2}(k\cdot\alpha_{-1}%
)(\alpha_{-1}\cdot\alpha_{-1})+9(\alpha_{-1}\cdot\alpha_{-2})\nonumber\\
&  +21(k\cdot\alpha_{-1})(k\cdot\alpha_{-2})+25(k\cdot\alpha_{-3})]\left\vert
0,k\right\rangle , \label{41}%
\end{align}
a symmetric type I spin two ZNS%

\begin{equation}
\lbrack2\theta_{\mu\nu}\alpha_{-1}^{(\mu}\alpha_{-2}^{\nu)}+k_{\lambda}%
\theta_{\mu\nu}\alpha_{-1}^{\lambda\mu\nu}]\left\vert 0,k\right\rangle
,k\cdot\theta=\eta^{\mu\nu}\theta_{\mu\nu}=0,\theta_{\mu\nu}=\theta_{\nu\mu}
\label{42}%
\end{equation}
where $\alpha_{-1}^{\lambda\mu\nu}\equiv\alpha_{-1}^{\lambda}\alpha_{-1}^{\mu
}\alpha_{-1}^{\nu}$ and two vector ZNS%
\begin{align}
\lbrack(\frac{5}{2}k_{\mu}k_{\nu}\theta_{\lambda}^{\prime}+\eta_{\mu\nu}%
\theta_{\lambda}^{\prime})\mathcal{\alpha}_{-1}^{(\mu\nu\lambda)}+9k_{\mu
}\theta_{\nu}^{\prime}\mathcal{\alpha}_{-1}^{(\mu\nu)}+6\theta_{\mu}^{\prime
}\mathcal{\alpha}_{-1}^{\mu}]\left\vert 0,k\right\rangle ,\theta\cdot k  &
=0,\label{43}\\
\lbrack(\frac{1}{2}k_{\mu}k_{\nu}\theta_{\lambda}+2\eta_{\mu\nu}%
\theta_{\lambda})\mathcal{\alpha}_{-1}^{(\mu\nu\lambda)}+9k_{\mu}\theta_{\nu
}\mathcal{\alpha}_{-1}^{[\mu\nu]}-6\theta_{\mu}\mathcal{\alpha}_{-1}^{\mu
}]\left\vert 0,k\right\rangle ,\theta\cdot k  &  =0. \label{44}%
\end{align}
Note that Eq.(\ref{43}) and Eq.(\ref{44}) are linear combinations of a type I
and a type II ZNS. This completes the four ZNS at the second massive level
$M^{2}=$ $4$.

In the Regge limit, the scalar ZNS in Eq.(\ref{41}) gives the RZNS%
\begin{equation}
\lbrack25(\alpha_{-1}^{P})^{3}+9\alpha_{-1}^{P}(\alpha_{-1}^{L})^{2}%
+9\alpha_{-1}^{P}(\alpha_{-1}^{T})^{2}-9\alpha_{-2}^{L}\alpha_{-1}^{L}%
-9\alpha_{-2}^{T}\alpha_{-1}^{T}-75\alpha_{-2}^{P}\alpha_{-1}^{P}%
+50\alpha_{-3}^{P}]\left\vert 0,k\right\rangle . \label{R4.1}%
\end{equation}
For the type I spin two ZNS in Eq.(\ref{42}), we define $\theta_{\mu\nu}%
=\sum_{\alpha,\beta}e_{\mu}^{\alpha}e_{\nu}^{\beta}u_{\alpha\beta}$, symmetric
and transverse conditions on $\theta_{\mu\nu}$ implies%
\begin{equation}
u_{\alpha\beta}=u_{\beta\alpha};u_{PP}=u_{PL}=u_{PT}=0. \label{MM}%
\end{equation}
Naively, the traceless condition on $\theta_{\mu\nu}$ implies%
\begin{equation}
u_{PP}-u_{LL}-u_{TT}=0. \label{Naive}%
\end{equation}
However, for the reason which will become clear later that one needs to
include at least one component $u_{NN}$ perpendicular to the scattering plane
and modify Eq.(\ref{Naive}) to%
\begin{equation}
u_{PP}-u_{LL}-u_{TT}-u_{NN}=0. \label{NN}%
\end{equation}
Note that, in the Regge limit, Eq.(\ref{NN}) reduces to Eq.(\ref{Naive}).
However, the solutions for Eq.(\ref{MM}) and Eq.(\ref{NN}) give three RZNS%
\begin{equation}
(\alpha_{-1}^{L}\alpha_{-2}^{L}-\alpha_{-1}^{P}\alpha_{-1}^{L}\alpha_{-1}%
^{L})|0,k\rangle, \label{R4.2}%
\end{equation}%
\begin{equation}
(\alpha_{-1}^{T}\alpha_{-2}^{T}-\alpha_{-1}^{P}\alpha_{-1}^{T}\alpha_{-1}%
^{T})|0,k\rangle, \label{R4.3}%
\end{equation}%
\begin{equation}
(\alpha_{-1}^{(L}\alpha_{-2}^{T)}-\alpha_{-1}^{P}\alpha_{-1}^{L}\alpha
_{-1}^{T})|0,k\rangle, \label{R4.4}%
\end{equation}
while Eq.(\ref{MM}) and Eq.(\ref{Naive}) give only two RZNS%
\begin{equation}
(\alpha_{-1}^{L}\alpha_{-2}^{L}-\alpha_{-1}^{P}\alpha_{-1}^{L}\alpha_{-1}%
^{L}-\alpha_{-1}^{T}\alpha_{-2}^{T}+\alpha_{-1}^{P}\alpha_{-1}^{T}\alpha
_{-1}^{T})|0,k\rangle\label{R4.23}%
\end{equation}
and Eq.(\ref{R4.4}). Note that Eq.(\ref{R4.23}) is just a linear combination
of Eq.(\ref{R4.2}) and Eq.(\ref{R4.3}). For the high-energy fixed angle
calculation in \cite{ChanLee1,ChanLee2}, the corresponding extra ZNS will not
affect the final result there. The vector ZNS in Eq.(\ref{43}) gives two RZNS%
\begin{equation}
\lbrack6\alpha_{-3}^{T}-18\alpha_{-1}^{(P}\alpha_{-2}^{T)}+9\alpha_{-1}%
^{P}\alpha_{-1}^{P}\alpha_{-1}^{T}+\alpha_{-1}^{L}\alpha_{-1}^{L}\alpha
_{-1}^{T}+\alpha_{-1}^{T}\alpha_{-1}^{T}\alpha_{-1}^{T}]|0,k\rangle,
\label{R4.5}%
\end{equation}%
\begin{equation}
\lbrack6\alpha_{-3}^{L}-18\alpha_{-1}^{(P}\alpha_{-2}^{L)}+9\alpha_{-1}%
^{P}\alpha_{-1}^{P}\alpha_{-1}^{L}+\alpha_{-1}^{L}\alpha_{-1}^{L}\alpha
_{-1}^{L}+\alpha_{-1}^{L}\alpha_{-1}^{T}\alpha_{-1}^{T}]|0,k\rangle.
\label{R4.6}%
\end{equation}
\ The vector ZNS in Eq.(\ref{44}) gives two RZNS%
\begin{equation}
\lbrack3\alpha_{-3}^{T}+9\alpha_{-1}^{[P}\alpha_{-2}^{T]}-\alpha_{-1}%
^{L}\alpha_{-1}^{L}\alpha_{-1}^{T}-\alpha_{-1}^{T}\alpha_{-1}^{T}\alpha
_{-1}^{T}]|0,k\rangle, \label{R4.7}%
\end{equation}%
\begin{equation}
\lbrack3\alpha_{-3}^{L}+9\alpha_{-1}^{[P}\alpha_{-2}^{L]}-\alpha_{-1}%
^{L}\alpha_{-1}^{L}\alpha_{-1}^{L}-\alpha_{-1}^{L}\alpha_{-1}^{T}\alpha
_{-1}^{T}]|0,k\rangle. \label{R4.8}%
\end{equation}
There are totally 8 RZNS at the mass level $M^{2}=$ $4$.

{\footnotesize \itemsep=0pt
\providecommand{\eprint}[2][]{\href{http://arxiv.org/abs/#2}{arXiv:#2}} }

\end{document}